\documentclass[prb,twocolumn,floatfix,superscriptaddress,amssymb,amsmath]{revtex4}
\usepackage{graphicx}
\usepackage{amssymb}

\begin{document}
\title{Spin relaxation rates in quasi-one-dimensional coupled
quantum dots}
\author{C.\ L.\ Romano\footnote{carlu@df.uba.ar}}
\affiliation{Department of Physics ``J.\ J.\ Giambiagi",
University of Buenos Aires, Ciudad Universitaria, Pab.\ I,
C1428EHA Buenos Aires, Argentina}
\author{P.\ I.\ Tamborenea}
\affiliation{Department of Physics ``J.\ J.\ Giambiagi",
University of Buenos Aires, Ciudad Universitaria, Pab.\ I,
C1428EHA Buenos Aires, Argentina}
\author{S.\ E.\ Ulloa}
\affiliation{Department of Physics and Astronomy, and
Nanoscale and Quantum Phenomena Institute, Ohio University,
Athens, Ohio 45701-2979}
\date{\today}

\begin{abstract}
We study theoretically the spin relaxation rate in
quasi-one-dimensional coupled double semiconductor quantum dots.
We consider InSb and GaAs-based systems in the presence of the
Rashba spin-orbit interaction, which causes mixing of
opposite-spin states, and allows phonon-mediated transitions
between energy eigenstates. Contributions from all phonon modes
and coupling mechanisms in zincblende semiconductors are taken
into account. The spin relaxation rate is shown to display a
sharp, cusp-like maximum as function of the interdot-barrier
width, at a value of the width which can be controlled by an
external magnetic field. This remarkable behavior is associated
with the symmetric-antisymmetric level splitting in the structure.
\end{abstract}

\maketitle

Spin-related phenomena in semiconductors attract much attention as
they are the foundation of the emerging fields of
\textit{spintronics} \cite{zut-fab-das} and \textit{quantum
computing} in semiconductor systems.\cite{los-div} Quantum dots
(QD) are particularly promising since they offer relatively long
spin coherence times, a key requirement in quantum information
processing. Electron spin relaxation in QDs has been studied
recently theoretically \cite{kha-naz,erl-naz-fal,mer-efr-ros,%
tah-fri-joy,woo-rei-lya,bur-los,sou-das,sem-kim-03,lev-ras,%
kha-los-gla,gla-kim,dic-haw,che-wu-lu,sem-kim-04,gol-kha-los,%
aba-mar,tsi-loz-gog,coi-los,cha-mal-cha,bul-los,mar-aba,des-ull}
and experimentally.
\cite{gup-aws-pen-ali,fuj-aus-tok-nature,fuj-aus-tok-PRL,han-wit-nav,%
tac-oht-yam} In this letter, we study spin relaxation rates in
coupled double QDs, a type of structure that offers a very useful
control parameter, i.e.\ the interdot separation, or barrier
width. In particular, we consider here quasi-one-dimensional
(quasi-1D) dot structures produced in {\em nanowhiskers} or {\em
nanorods} studied experimentally.  We have recently studied the
electronic states of such quasi-1D double dots,\cite{rom-ull-tam}
and analyzed the spin-mixed character that arises from the
spin-orbit (\textbf{SO}) interaction. These structures can be
designed so that only the Rashba SO interaction (enabled by {\it
structural} asymmetry) is present.\cite{rom-ull-tam} In this
paper, we calculate rates of spin transitions induced by phonon
scattering between Rashba spin-mixed states, taking into account
the different acoustic-phonon modes present in zincblende
semiconductors. We find that the spin relaxation rate shows a
strong dependence on the interdot-barrier width and can,
furthermore, be tuned with an external magnetic field.  This
provides interesting flexibility in the control of electronic spin
states.

Let us denote by \textit{z} the coordinate in the longitudinal
direction of the quasi-1D coupled double quantum-dot structure.
$V_{z}(z)$ is the confining potential that defines the pair of
coupled QDs. We assume a narrow whisker width of 2-5 nm, that each
dot has a length of $30\,\mbox{nm}$, and take the width of the
barrier between dots as a variable parameter. For the case of
Rashba interaction,\cite{rom-ull-tam,ras,byc-ras} and introducing
a weak magnetic field in the $z$ direction that breaks the spin
degeneracy (but produces no diamagnetic shift), the Hamiltonian
takes the form\cite{rom-ull-tam}
\begin{equation}
H=\frac{\textbf{P}^{2}}{2m^{*}}+H_R+\frac{1}{2}\textit{g}
\mu_{B}B\sigma_{z},
\end{equation}
where $\textbf{P}=(P_x, P_y,P_z)$, and
$\sigma=(\sigma_x,\sigma_y,\sigma_z)$ is the Pauli matrix
vector.\cite{footnote1p} After taking average $\langle ...
\rangle$ over the ground states of the lateral-confining
potentials, the SO term becomes\cite{footnote1,rom-ull-tam}
\begin{eqnarray}
H_{1dR} = \frac{\gamma_{R}}{\hbar}p_{z}
\left\langle\frac{\partial V_x}{\partial x}\right\rangle
\left( \sigma_x - \sigma_y \right),
\end{eqnarray}
where $\gamma_{R}$ is material specific.

We diagonalize the Hamiltonian to take full account of the SO
effects, which result in spin mixing of eigenstates.  We calculate
relaxation rates due to acoustic-phonon scattering between the
ground state and the next two energy eigenstates in InSb and GaAs
QDs via Fermi's Golden Rule:
\begin{equation}
\Gamma_{i \rightarrow f} = \frac{2\pi}{\hbar}
\sum_{\mathbf{Q},\alpha} |\langle f| U_{Q,\alpha} |i\rangle|^{2}
\emph{n}\, \delta(\Delta E - \hbar \omega_{\alpha}),
\end{equation}
where $\mathbf{Q}=(q_x,q_y,q_z)=(\mathbf{q},q_z)$ is the phonon
momentum; $\alpha$ indicates the acoustic phonons, and includes
longitudinal, $\ell$, and transverse, $t=$ TA1 and TA2 modes;
$\Delta E = E_{f}-E_{i}$; and $n$ is the Bose-Einstein phonon
distribution with energy $\hbar\omega_{\alpha}=\hbar
\textit{c}_{\alpha}Q$. $\langle f|$, $\langle i|$ are the final
and initial states obtained by exact (numerical) diagonalization
of the Hamiltonian. The potential $U_{q,\alpha}$ includes both
deformation $\Xi_{\ell}(\mathbf{Q}) $ and piezoelectric
$\Lambda_{\ell}(\mathbf{Q})$ contributions:\cite{mah}
\begin{equation}
U_{q,\alpha=\ell,t}=(\Xi_{\ell}(\mathbf{Q})+
i\Lambda_{\ell,t}(\mathbf{Q})) \, e^{i \mathbf{Q} \cdot
\mathbf{r}}.
\end{equation}
For zincblende semiconductors, the phonon potentials read (in
cylindrical coordinates):\cite{des-ull}
$\Xi_{\ell}(\textbf{Q}) =
   \Xi_0 \sqrt{\frac{\hbar}{2DVc_{\ell}}} \sqrt{Q}$,
$\Lambda_{\ell}(\textbf{Q}) =
   \frac{6 \pi e h_{14}}{\kappa}
   \sqrt{\frac{\hbar}{2 D V c_{\ell}}}
    \sin(2\phi)
   \frac{q^{2}q_{z}}{Q^{7/2}}$,
$\Lambda_{\mathrm{TA1}}(\textbf{Q}) =
   \frac{4 \pi e h_{14}}{\kappa}
   \sqrt{\frac{\hbar}{2 D V c_{\mathrm{TA}}}}
    \cos(2\phi)
   \frac{qq_{z}}{Q^{5/2}}$
and
$\Lambda_{\mathrm{TA2}}(\textbf{Q}) =
   \frac{2 \pi e h_{14}}{\kappa}
   \sqrt{\frac{\hbar}{2 D V c_{\mathrm{TA}}}}
    \sin(2\phi)
   \frac{q^3}{Q^{7/2}}(2 \frac{q_z^2}{q^2}-1)$,
where $\Xi_0$ and $e h_{14}$ are the bulk constants, $\kappa$ is
the dielectric constant, $D$ is the mass density, $V$ is the volume,
and $c_{\ell}$,
$c_{\mathrm{TA1}} = c_{\mathrm{TA2}} = c_{\mathrm{TA}}$ are the sound
velocities for each mode.


\begin{figure}[hbp]
\includegraphics*[width=9cm]{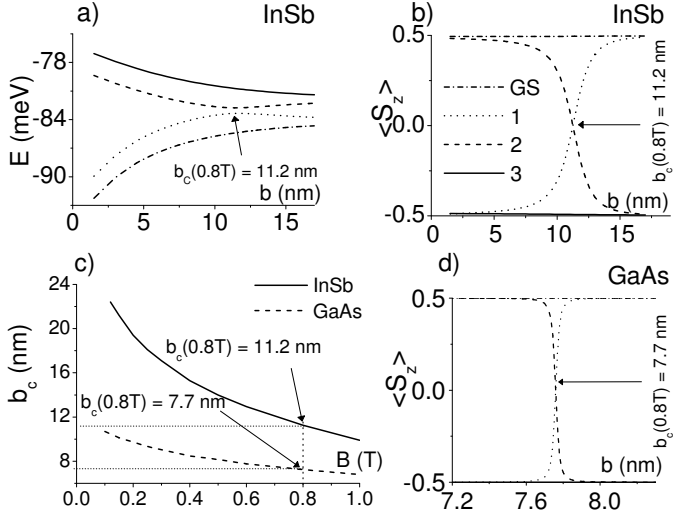}
\caption{
 Energy levels (a) and mean value of $S_z$ (b) for the four
 lowest-energy eigenstates as function of the width of the central
 barrier, $b$, for InSb at $B =0.8\,\mbox{T}$.
 Notice states 1 and 2 anticross and switch spin at $b_c(B)$.
 (c) shows the crossing barrier $b_c$ {\it vs} magnetic field
 for GaAs and InSb QDs. Panel (d) is the same as (b) but for GaAs.
}
\label{fig1}
\end{figure}

For all the calculations reported here we assume a temperature of
$298\,\mbox{K}$ and a Rashba structural parameter, $\left\langle
\frac{\partial V_x}{\partial x}\right\rangle =
5\,\mbox{meV/nm}$.\cite{footnote2}
In Fig.\ 1(a) (and 1(b)) the four lowest energy levels (and mean
value of $S_z$) are shown as function of the width of the central
barrier, $b$, for a magnetic field $B=0.8\,\mbox{T}$, and fixed
dot size. We note that levels 1 and 2 show an anticrossing at a
barrier width value where the $S_z$ values switch,
$b_c(0.8\,\mbox{T})=11.2\,\mbox{nm}$, seen in Fig.\ 1(b). At low
$b$-values, state 1 is the spatially-symmetric double-dot state
with spin down, while state 2 is the spatially-antisymmetric state
with spin up.  Increasing barrier width decreases the
symmetric-antisymmetric splitting, allowing SO to produce strong
mixing, which results in the anticrossing and spin-switching we
see in Fig.\ 1.  This ``crossing value" is magnetic field
dependent, as shown in Fig.\ 1(c) for InSb and GaAs, and shows a
monotonic drop with increasing $B$, as one would expect.

Figure 2 shows the contributions to the spin relaxation
(\textbf{SR}) rate in InSb and GaAs coupled QDs due to the four
different acoustic-phonon couplings: deformation, and
piezoelectric longitudinal (LA), and transverse (TA1 and TA2)
potentials. The rates shown correspond to the transition between
the two lowest-energy states (as in Fig.\ 1(a)). In InSb QDs, we
note that the SR rate is dominated by the deformation potential
while in GaAs, it is dominated by the piezoelectric TA1
potential.\cite{des-ull} Notice that in InSb QDs, the
contributions from the piezo TA2 and LA are several orders of
magnitude weaker than the others for magnetic fields beyond $\sim
0.3\,\mbox{T}$. The same is true of the deformation potential in
comparison to the other mechanisms in GaAs. The strong SO
interaction (smaller bandgap) in InSb, results in much higher SR
rates for that material than for GaAs, as one can observe in Fig.\
2.


\begin{figure}[tbp]
\includegraphics*[width=8cm]{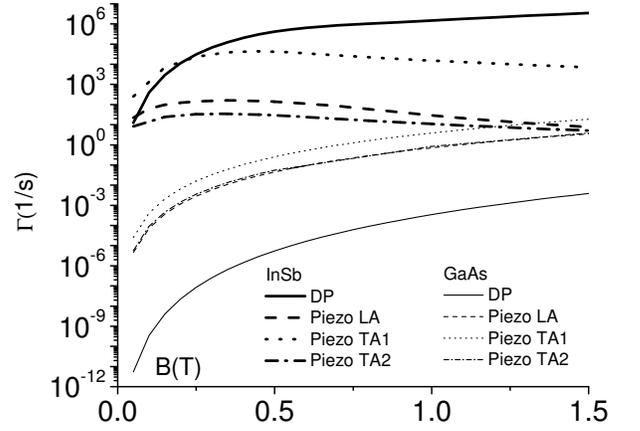}
\caption{Spin relaxation rates from different acoustic-phonon
potentials: deformation (DP), piezoelectric longitudinal (Piezo
LA), and transverse 1 (Piezo TA1) and 2 (Piezo TA2) potentials,
for InSb (thick lines, upper four curves) and GaAs (thin lines)
QDs as function of applied magnetic field. Barrier width $b=
30$nm. } \label{fig2}
\end{figure}


\begin{figure}[hbp]
\includegraphics*[width=8cm]{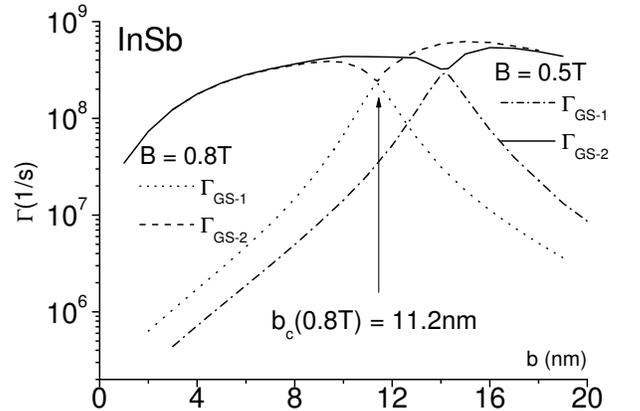}
\caption{Spin relaxation rate as function of central barrier width
in InSb QDs, for two different values of the magnetic field ($B=
0.5 \,\mathrm{and}\, 0.8\, \mbox{T}$), and different transitions:
from ground to first excited state (GS-1), and to second excited
state (GS-2).} \label{fig3}
\end{figure}


\begin{figure}[tbt]
\includegraphics*[width=7cm]{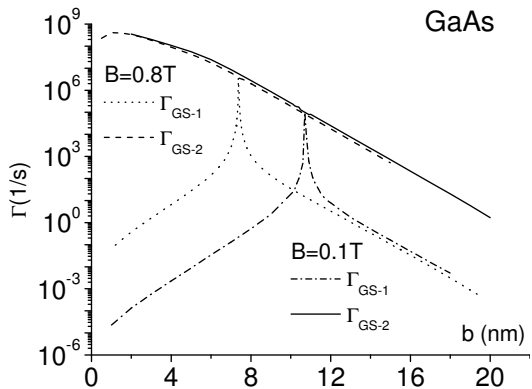}
\caption{Same as Fig.\ 3 but for GaAs at $B= 0.1 \,\mathrm{and}\,
0.8\, \mbox{T}$.} \label{fig4}
\end{figure}

Figure \ref{fig3} shows the \emph{total} transition rate from the
ground state to the first excited state (GS-1) and to the second
excited state (GS-2) for InSb QDs and two values of the magnetic
field, $B=0.5$ and 0.8 T. We see that for a given field, the rate
GS-1 shows a cusp at a barrier width that coincides with the
crossing width, $b_c$, as introduced in Fig.\ \ref{fig1}.
Accordingly, the cusp position in $\Gamma$ shifts down with
increasing magnetic field. An analogous situation can be seen in
Fig.\ \ref{fig4} for GaAs QDs, with $B=0.1$ and 0.8 T. Here the
cusp is sharper than for InSb, in correspondence with the weaker
SO, as shown in Figs.\ 1(b) and 1(d). It must be noted that the
transition GS-1 is not strictly speaking a ``spin-flip"
transition, since the states involved are not spin eigenstates.
Indeed, the GS is practically a pure spin-up state (Fig.\ 1(b))
but the state 1 has a strong spin admixture, and switches from
being mostly spin-down to mostly spin-up as a function of the
barrier width, $b$. The middle point in this transformation is
given by $b_c$, where the cusp in the transition rate occurs. The
$\Gamma_{GS-2}$ values, on the other hand, show a slight dip at
the same $b_c$ position, reflecting also the mixed spin character
of the 2 state.  The appearance of the cusp on the lowest-energy
SR rates clearly arises from the enhanced level mixing when the
symmetric-antisymmetric states anticross in the double QD.

To summarize, we have studied the phonon-mediated electronic
transitions between (Rashba-SO) spin-mixed states in GaAs and InSb
coupled QDs. This rate shows a cusp-like maximum as a function of
the separation between the dots. The position of this maximum can
be controlled with a small external magnetic field, and can in
principle be used to significantly change the electronic SR rates
in the system.

We thank C.F. Destefani, R. Romo, C. B\"usser, and M. Zarea for
useful discussions, and support from CMSS at OU, UBACyT 2004-2007,
ANPCyT 03-11609, and NSF-CONICET through a CIAM-NSF collaboration
0336431. P.I.T.\ is a researcher of CONICET.


\begin{thebibliography}{99}

\bibitem{zut-fab-das} I.\ Zutic, J.\ Fabian, and S.\ Das Sarma,
Rev.\ Mod.\ Phys.\ {\bf 76}, 323 (2004).

\bibitem{los-div} D.\ Loss and D.\ P.\ DiVincenzo,
Phy.\ Rev.\ A {\bf 57}, 120 (1998).

\bibitem{kha-naz} A.\ V.\ Khaetskii and Y.\ V.\ Nazarov,
Phys.\ Rev.\ B {\bf 61}, 12639 (2000); {\it ibid.} {\bf 64}, 125316 (2001).

\bibitem{erl-naz-fal} S.\ I.\ Erlingsson, Yu.\ V.\ Nazarov, and V.\ I.\ Fal'ko,
Phys. Rev. B {\bf 64}, 195306 (2001).

\bibitem{mer-efr-ros} I.\ A.\ Merkulov, Al.\ L.\ Efros, and M.\ Rosen,
Phys.\ Rev.\ B {\bf 65}, 205309 (2002).

\bibitem{tah-fri-joy} C.\ Tahan, M.\ Friesen, and R.\ Joynt,
Phys.\ Rev.\ B {\bf 66}, 035314 (2002).

\bibitem{woo-rei-lya} L.\ M.\ Woods, T.\ L.\ Reinecke, and Y.\ Lyanda-Geller,
Phys.\ Rev.\ B {\bf 66}, 161318 (2002).

\bibitem{bur-los} G.\ Burkard and D.\ Loss, in {\it Semiconductor
Spintronics and Quantum Computation}, edited by D.\ D.\ Awschalom,
D.\ Loss, and N.\ Samarth (Springer, New York, 2002).

\bibitem{sou-das} R.\ de Sousa and S.\ Das Sarma,
Phys.\ Rev.\ B {\bf 67}, 033301 (2003).

\bibitem{sem-kim-03} Y.\ G.\ Semenov and K.\ W.\ Kim,
Phys.\ Rev.\ B {\bf 67}, 073301 (2003).

\bibitem{lev-ras} L.\ S.\ Levitov and E.\ I.\ Rashba,
Phys.\ Rev.\ B {\bf 67}, 115324 (2003).

\bibitem{kha-los-gla} A.\ Khaetskii, D.\ Loss, and L.\ Glazman,
Phys.\ Rev.\ B {\bf 67}, 195329 (2003).

\bibitem{gla-kim} B.\ A.\ Glavin and K.\ W.\ Kim,
Phys.\ Rev.\ B {\bf 68}, 045308 (2003).

\bibitem{dic-haw} S.\ Dickmann and P.\ Hawrylak,
JETP Lett.\ {\bf 77}, 30 (2003).

\bibitem{che-wu-lu} J.\ L.\ Cheng, M.\ W.\ Wu, and C.\ L\"u,
Phys.\ Rev.\ B {\bf 69}, 115318 (2004).

\bibitem{sem-kim-04} Y.\ G.\ Semenov and K.\ W.\ Kim,
Phys.\ Rev.\ Lett.\ {\bf 92}, 026601 (2004).

\bibitem{gol-kha-los} V.\ N.\ Golovach, A.\ Khaetskii, and D.\ Loss,
Phys.\ Rev.\ Lett.\ {\bf 93}, 016601 (2004).

\bibitem{aba-mar} V.\ A.\ Abalmassov and F.\ Marquardt,
Phys. Rev. B {\bf 70}, 075313 (2004).

\bibitem{tsi-loz-gog} E.\ Tsitsishvili, G.\ S.\ Lozano, and A.\ O.\ Gogolin,
Phys. Rev. B {\bf 70}, 115316 (2004).

\bibitem{coi-los} W.\ A.\ Coish and D.\ Loss,
Phys.\ Rev.\ B {\bf 70}, 195340 (2004).

\bibitem{cha-mal-cha} C.-H. Chang, A.\ G.\ Mal'shukov, and K.\ A.\ Chao,
Phys.\ Rev.\ B {\bf 70}, 245309 (2004).

\bibitem{bul-los} D.\ V.\ Bulaev and D.\ Loss, Phys.\ Rev.\ B {\bf 71},
205324 (2005).

\bibitem{mar-aba}  F.\ Marquardt and V.\ A.\ Abalmassov,
Phys. Rev. B {\bf 71}, 165325 (2005).

\bibitem{des-ull} C.\ F.\ Destefani and S.\ E.\ Ulloa, cond-mat/0412520.

\bibitem{gup-aws-pen-ali} J.\ A.\ Gupta, D.\ D.\ Awschalom, X.\ Peng,
and A.\ P.\ Alivisatos,
Phys.\ Rev.\ B {\bf 59}, R10421 (1999).

\bibitem{fuj-aus-tok-nature}
T.\ Fujisawa, D.\ G.\ Austing, Y.\ Tokura, Y.\ Hirayama, and S.\ Tarucha,
Nature {\bf 419}, 278 (2002).

\bibitem{fuj-aus-tok-PRL} T.\ Fujisawa, D.\ G.\ Austing, Y.\ Tokura,
Y.\ Hirayama, and S.\ Tarucha
Phys.\ Rev.\ Lett.\ {\bf 88}, 236802 (2002).

\bibitem{han-wit-nav} R.\ Hanson, B.\ Witkamp, L.\ M.\ K.\ Vandersypen,
L.\ H.\ Willems van Beveren, J.\ M.\ Elzerman, and L.\ P.\ Kouwenhoven,
Phys.\ Rev.\ Lett.\ {\bf 91}, 196802 (2003).

\bibitem{tac-oht-yam} A.\ Tackeuchi, R.\ Ohtsubo, K.\ Yamaguchi, M.\ Murayama,
T.\ Kitamura, T.\ Kuroda, and T.\ Takagahara, Appl.\ Phys.\ Lett.\ {\bf 84},
3576 (2004).

\bibitem{rom-ull-tam} C.\ L.\ Romano, S.\ E.\ Ulloa, and P.\ I.\ Tamborenea,
Phys.\ Rev.\ B {\bf 71}, 035336 (2005).

\bibitem{ras} E.\ I.\ Rashba, Sov.\ Phys.\ Solid State {\bf 2}, 1109 (1960).

\bibitem{byc-ras} Y.\ A.\ Bychkov and E.\ I.\ Rashba, JETP Lett.\
{\bf 39}, 78 (1984);
J.\ Phys. C {\bf 17}, 6039 (1984).

\bibitem{footnote1p}The strong $xy$-confinement of the nanowhisker
suppresses diamagnetic effects for the weak $B$-fields we consider
here.

\bibitem{footnote1} We consider confining potentials $V_x(x)=V_y(y)$ without
inversion symmetry, a situation that also cancels the Dresselhaus
SO terms (see Ref.\ \onlinecite{rom-ull-tam}).

\bibitem{mah} G.\ D.\ Mahan, {\em Many Particle Physics} (Plenum, New York, 1981).

\bibitem{footnote2} We will report the dependence of the relaxation rates on
these parameters in a separate publication.

\end{thebibliography}
\end{document}